# LANGMUIR

Article

# Entrapment and Dissolution of Microbubbles Inside Microwells

Xiaolai Li,[†,§] Yuliang Wang,[*,†,‡] Binglin Zeng,[†] Yanshen Li,[§] Huanshu Tan,[§] Harold J. W. Zandvliet,[∥] Xuehua Zhang,[*,⊥,§] and Detlef Lohse[*,§]

[†]School of Mechanical Engineering and Automation and [‡]Beijing Advanced Innovation Center for Biomedical Engineering, Beihang University, 37 Xueyuan Road, Haidian District, Beijing 100191, China

[§]Physics of Fluids Group, Department of Applied Physics, J. M. Burgers Centre for Fluid Dynamics and [∥]Physics of Interfaces and Nanomaterials, MESA+ Institute for Nanotechnology, University of Twente, P.O. Box 217, Enschede 7500 AE, The Netherlands

[⊥]Department of Chemical and Materials Engineering, University of Alberta, 12-211 Donadeo Innovation Centre for Engineering, Edmonton, Alberta, Canada T6G1H9

**ABSTRACT:** The formation and evolution of immersed surface micro- and nanobubbles are essential in various practical applications, such as the usage of superhydrophobic materials, drug delivery, and mineral flotation. In this work, we investigate the entrapment of microbubbles on a hydrophobic surface, structured with microwells, when water flow passes along, and the subsequent microbubble dissolution. At entrapment, the microbubble is initially pinned at the edge of the microwell. At some point, the three-phase contact line detaches from one side of the edge and separates from the wall, after which it further recedes. We systematically investigate the evolution of the footprint diameter and the contact angle of the entrapped microbubbles, which reveals that the dissolution process is in the constant contact angle mode. By varying the gas undersaturation level, we quantify how a high gas undersaturation enhances the dissolution process, and compare with simplified theoretical predictions for dissolving bubbles on a plane surface. We find that geometric partial blockage effects of the diffusive flux out of the microbubble trapped in the microwell lead to reduced dissolution rates.

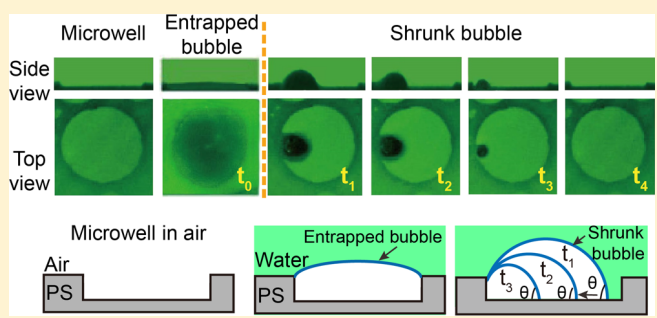

## ■ INTRODUCTION

Submicron surface bubbles, namely nanobubbles at solid−liquid interfaces with heights between 5 and 100 nm and footprint diameters between 50 and 800 nm, have extensively been studied over the last 2 decades.[1] The surface bubbles play an important role in various chemical and physical processes and have numerous potential applications,[2] such as mineral flotation and separation,[3] transport in nanofluidic devices,[4] nanostructured surface fabrication[5,6] or application in the context of catalysis and electrolysis.[7,8] So far, various theoretical and experimental studies have been performed to investigate surface nanobubbles[9,10] and their intriguing properties, such as their stability, their small contact angle, and collective effects.[1,11−14] Surface nanobubbles can be obtained by several methods,[1] such as solvent exchange,[9] spontaneous generation at immersion,[10] electrochemical or catalytic production,[15] etc. However, it is still challenging to achieve full control over the formation of surface nanobubbles.

To achieve such controllable generation of nanobubbles, it is essential to understand the formation mechanism. The so-called crevice model has been proposed to explain the formation mechanism of nucleating surface micro and nanobubbles, e.g., under pressure reduction.[16,17] According to this model, gas is entrapped in crevices on surfaces and forms bubble nuclei. These bubble nuclei can then grow by diffusion of gas from the surrounding oversaturated liquid or due to expansion upon the reduction of the liquid pressure. Numerous experimental studies have been conducted at a microscale to validate this nucleation model. For example, techniques such as centrifugation, shock wave, and acoustics were applied to drastically reduce the liquid pressure to a negative value to find the cavitation pressure threshold.[18−21] These studies mostly focused on controlling the liquid condition after gas entrapment. However, the detailed entrapment process, namely the entrapment dynamics, in which the surface structures play an important role because they affect liquid flows and provide pinning sites, was less studied.

Several previous studies have used structured surfaces to explore the nucleation mechanism and controllable interfacial nanobubble formation.[22−25] With nanopatterned hydrophobic/hydrophilic surfaces, Agrawal et al.[26] found that interfacial nanobubbles only nucleate on hydrophobic domains. In our previous work,[24] the in situ entrapment of nanobubbles was observed with an atomic force microscope (AFM). The size of the nucleated surface nanobubbles is linearly correlated with that of the surface nanopores on hydrophobic surfaces, which supports that the nanobubbles form from gas entrapment on the nanoscopic cavities. Although the entrapped nanobubbles have been observed in these works, further investigation on the entrapment dynamics is extremely challenging due to the limited temporal resolution of the measurement equipment at nanoscale, e.g., AFM. Essentially, the entrapments of surface microbubbles and surface nanobubbles share the same mechanism. Therefore, the investigation of nucleation dynamics of surface microbubbles, which is experimentally feasible to be visualized, will lead us to a better understanding of the nanobubble entrapment dynamics.

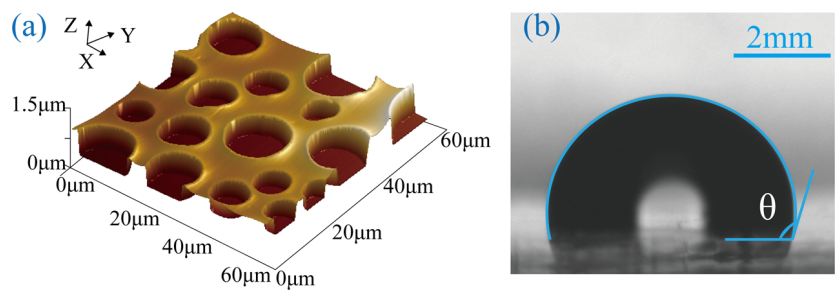

**Figure 1.** (a) Characterization of the structured PS surface. Three-dimensional morphology of the sample surface obtained with AFM. Microwells with different sizes were obtained on the surface. (b) A side view image of a sessile drop on the sample surface, indicating a water contact angle of $\theta$ = 109.

Recently, Langley et al.[27] demonstrated the entrapment-based microbubble formation and observed the dynamical process in their work. They studied air entrapment by performing drop impact experiment on nanoparticle-decorated surfaces. As the drops approach the solid surfaces, a central air disc is entrapped due to the deformation of the drop by the intervening air layer. After the drops touch the sample surface, the liquid blocks the escape path for the gas, leading to the entrapment of microbubbles in surface structures. The size of the microbubbles depends on the lateral roughness variation. These results confirm that bubble nucleation is mediated by gas entrapment and shed light on the dynamics of nucleation.

In this study, we aim to reveal entrapment and diffusive dynamics of surface microbubbles. Our work can provide (i) a better understanding of the surface bubble nucleation mechanism and (ii) a potential method of reproducible micro- and nanobubble formation. Here, we investigate the temporal evolution of the surface microbubbles entrapped by microwells on hydrophobic surfaces using a confocal microscope. To study the effect of the microwell diameter on entrapment and dissolution effectively, we prepared wells with different sizes on one sample. We will show that the microwells can be used to trap surface microbubbles. After this, the trapped surface microbubbles will dissolve. The detailed process of microbubble dissolution, as well as factors such as gas concentration and lateral size of surface microstructures on the microbubble dissolution will be investigated in detail.

## ■ EXPERIMENTAL SECTION

**Preparation of the Sample and Its Characterization.** Two solutions, polystyrene (PS)–toluene and water–acetone, were prepared to fabricate the structured polystyrene (PS) surface on a glass substrate. The PS–toluene solution was prepared by dissolving PS particles (molecular weigh 350 000, Sigma-Aldrich) into toluene (Mallinckrodt Chemical) with a concentration of 1.0% (weight). The water–acetone solution was made by mixing water and acetone with a water concentration of 5.0%. The two solutions were first mixed with a ratio of 1:3 (volume, PS–toluene to water–acetone). About 200 $\mu$L of the mixed solution was then dropped on a piece of glass substrate (20 mm × 20 mm). The deposited droplet solution will rapidly spread over the glass substrate. Due to the lower solubility in the acetone–water mixture compared to that in toluene, PS precipitates from the mixed solution and deposits on the glass substrate to form a film. Among the three different liquids (acetone, toluene, and water), acetone has the highest evaporation rate, followed by toluene. As a result, the water droplets remain on the PS film after acetone and toluene have evaporated within several seconds. Eventually, the remaining water droplets will also evaporate within about 10 min and the PS film with microwells (at the locations where the droplets were) remains.

The morphology of the surface was measured with an AFM (Resolve, Bruker) in the tapping mode, as shown in Figure 1a. The width of the microwells is in between 10 and 80 $\mu$m. We confirm that the bottom of the microwells is also coated by the PS film by scratching it with an AFM tip (NSC36/ALBS, MikroMasch). The thickness of the film at the bottom of the microwell is about 30 nm, whereas the film itself is in between 0.8 and 1.2 $\mu$m. It is noteworthy to point out that there are other surface preparation techniques available that are able to produce more monodisperse distributions of microwells. In this work it is of key importance to have microwells with different diameters as we want to study the effect of the bubble diameter on the entrapment and the dissolution dynamics of the bubble.

To estimate the surface hydrophobicity, a water drop of 5 $\mu$L was placed on the surface. The static contact angle of about 109° (on the droplet side) was obtained using a video-based optical contact angle measuring system (DataPhysics OCA15 Pro), see Figure 1b. This contact angle is slightly larger than that of around 95° obtained on continuous PS films.[28] Such surface microwells can be applied in many applications,[29–31] including serving as cell containers or scaffolds for cell growth,[32,33] microreactors for chemical reactions,[34] and nucleation sites for photonic crystals.[35,36]

**Water Deposition and Gas Concentration Control.** During the experiment, the prepared microstructured PS sample was clamped in a home designed microfluidic chamber. Then 3 mL of deionized water (Milli-Q Advantage A10 System, Germany) was deposited on the surface. This leads to air bubble entrapment in the microwells. The flow rates were controlled by a motorized syringe pump (Harvard; PHD 2000). The temporal evolution of the entrapped microsized surface bubbles was observed using a laser scanning confocal microscope (LSCM, Nikon Confocal Microscopes A1 system) with a 60× water immersion objective (CFI Apochromat 60XW NIR, numerical aperture = 1, working distance = 2.8 mm). To obtain the actual morphology of the structured surface, the sample was first immersed into fully degassed water (the gas concentration is 0.2 measured by an oxygen meter (Fibox 3 Trace, PreSens)), and subsequently was scanned with the confocal microscope.

To test the effect of the gas concentration on the microbubble dynamics, the experiment was conducted in both (nearly) air equilibrated water (AEW) and in partially degassed water (PDW). In the AEW experiment, a sample bottle containing water was kept open in air for 10 h. The measured gas concentration in the nearly air saturated water is 96.0%. For the PDW, the Milli-Q water was degassed for 3 min in a home-made vacuum chamber. After partial degassing, the measured air concentration was 63.0%. For visualization, water was labeled in yellow color with fluorescein isothiocyanate−dextran (Sigma-Aldrich, molecular weight, 70 000). During experiments, water flow rates from 0.5 to 3.0 mL/min were applied. All images were captured with the confocal microscope.

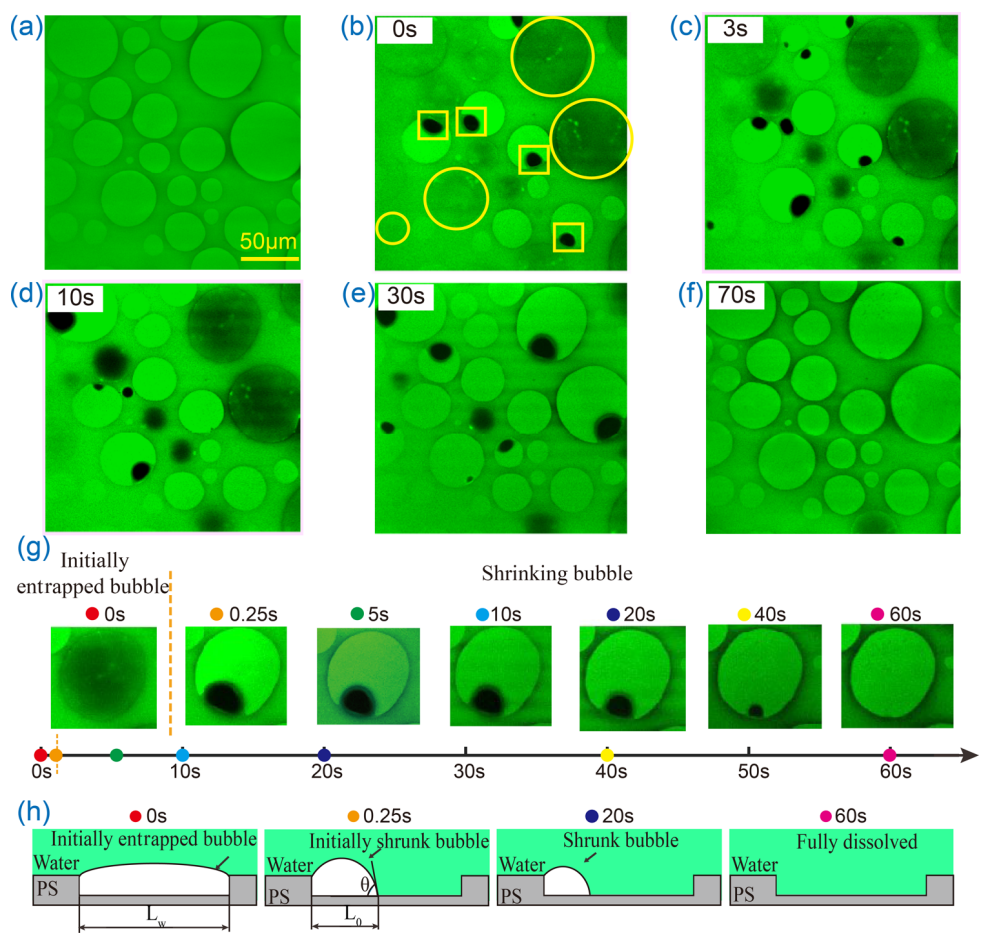

**Figure 2.** Evolution of the entrapped microbubbles in air equilibrated water. (a) The structured polystyrene surface captured in degassed water, showing the distribution of microwells on the surface. (b−f) Confocal microscopy images of the selected surface area at different times. The dark circular areas marked by yellow circles are the microwells covered by the initially entrapped microbubbles. The areas marked by yellow rectangles are the shrinking microbubbles. During shrinkage, the three-phase contact line keeps receding towards a pinned spot until the microbubble disappears completely. (g) The sequential images of an entrapped microbubble on a microwell. Initially, the microwell was completely filled by the entrapped microbubble. Then part of contact line detaches from the edge of the microwell and the formed microbubble gradually shrank and eventually disappeared. (h) Schematic illustration of the three phases of air entrapment in a microwell, fully entrapped, shrinking, and fully dissolved.

## ■ RESULTS AND DISCUSSION

**Entrapment of Surface Microbubbles.** When the flow front passes over the structured hydrophobic surface, air pockets are entrapped in the microwells.[37−39] The air entrapment occurs due to the large advancing contact angle in combination with the surface structure size, as explained by the crevice model.[37] The entrapped air then first remains in the microwells, as surface microbubbles are pinned to the edge of the surface cavities.

We subsequently observed the diffusive evolution of the entrapped surface microbubbles in the microwells. To obtain the actual topography of the structured surface before air entrapment, the sample was first immersed into fully degassed water. The bottom view of the microwell is shown in Figure 2a.

The circular areas with a green color indicate that no microbubble entrapment occurred in the fully degassed water (or the entrapped microbubbles dissolved immediately, see below). After this, the fully degassed water was removed.

The AEW was then injected into the microfluidic chamber. Subsequently, the sample was fully immersed in AEW, and we immediately captured the image of the entrapped microbubbles. At this stage, the microwells were fully filled with gas. After a certain time, part of the three-phase contact line detached from the edges of the microwells and the initially trapped microbubbles rapidly shrank, reducing their lateral sizes. Figure 2b depicts the area captured right after the sample was immersed into water. Due to the limited frame rate of the LSCM (4 fps was applied to achieve optimized imaging), the initial shrinkage for some of the microbubbles was not captured. As a result, some of the initially entrapped microbubbles shrank within the first frame. The areas marked with yellow circles correspond to the initially entrapped microbubbles, whereas the ones marked with yellow rectangles are the shrinking microbubbles.

Figure 2c−e show sequential images of the dissolving surface microbubbles. Noticeably, in Figure 2c, the bubbles appearing

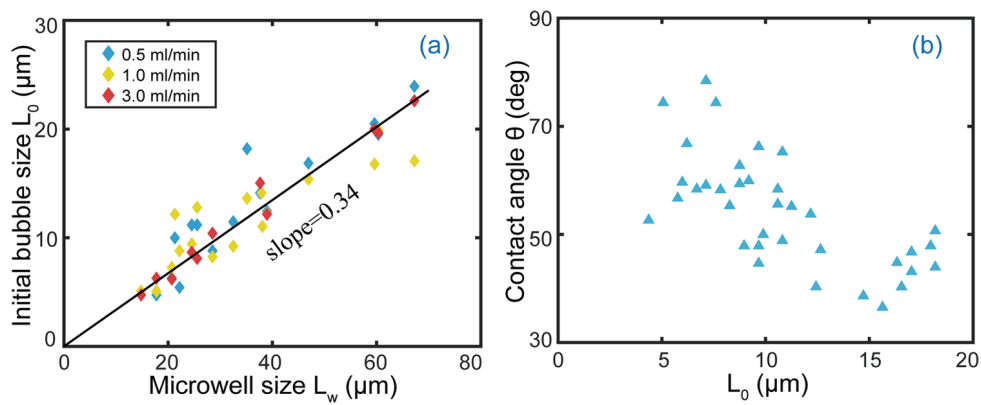

**Figure 3.** (a) Correlation of the lateral size $L_0$ of the initially shrunk microbubbles and the width $L_w$ of the microwells. The value $L_0$ approximately linearly increases with $L_w$. This implies that the larger well contains more entrapped gas. (b) Contact angle $\theta$ of the entrapped microbubbles as a function of $L_0$. The contact angle decreases with increasing $L_0$.

at the bottom left quadrant are formed from depinning of the previous entrapped bubbles, which are marked with yellow circles in Figure 2b. In addition, we observed that a bubble appears between two microwells (top left quarter). According to the sequential images between Figure 2b,c, we confirm that it jumps out of the well from its right side. In Figure 2f, all entrapped microbubbles have disappeared.

In Figure 2g, the sequential images of an initially entrapped microbubble and the corresponding shrinking microbubble in the microwell are presented. Initially, the three-phase contact line was pinned at the edge of the microwell and the complete well was covered by the air–water interface, as illustrated in Figure 2h (the first). After a certain period, the three-phase contact line detached from the edge of the well from one side and the microbubble rapidly shrank in size. The microbubbles right after the initial shrinking are referred to as initially shrunk microbubbles, as illustrated in Figure 2h (the second). Subsequently, the microbubble will gradually further shrink (Figure 2h (the third)) and eventually disappear (Figure 2h (the fourth)). The time required for the transition from the initially entrapped microbubbles to the initially shrunk microbubbles is much shorter than the following shrinkage period. Here, the frame which was taken right after the initial shrinkage is taken as the first frame for the analysis of the diffusive microbubble shrinkage dynamics.

Our results show that the lateral width $L_0$ of the initially shrunk microbubbles are related to the size of the microwells. Figure 3a depicts the correlation between $L_0$ and the microwell diameters $L_w$. One can see that $L_0$ linearly increases with $L_w$, with a slope of 0.34. This linear dependence is independent of the flow rates, at least for the applied flow rates of 0.5, 1.0, and 3.0 mL/min. This implies that larger microwells are able to trap more gas. The contact angles always measured on the gas side of the initially shrunk microbubbles are also size dependent. As shown in Figure 3b, the contact angle of the initially shrunk microbubbles changes from 30 to 80°, and slightly decrease with $L_0$.

**Dissolution of Entrapped Microbubbles.** From the above experimental observations, it is clear that the initially entrapped microbubbles will shrink and eventually completely disappear. However, the dynamics of the shrinking microbubbles still remains unknown. One example of microbubble shrinkage is shown in Figure 4a, in which the top and bottom row show the side and bottom view of the confocal microscopy images, respectively. From the side view images, one can see that both height and width of the microbubble gradually decrease with time until the microbubble eventually completely disappears. Moreover, images from both views indicate that one side of the three-phase contact line is pinned at the edge of the well and the other side keeps receding during the shrinking process. As an example, a schematic diagram of the microbubble shrinkage is shown in Figure 4b. The phenomenon that one side is pinned and the other side is detached is due to the heterogeneity of the pinning site, and thus, the pinning strength. The reason for such symmetry breaking is that the substrates in the experiments are never perfectly homogeneous. Slight differences in the surface properties are sufficient to lead to pinning or depinning on one side only.[40,41]

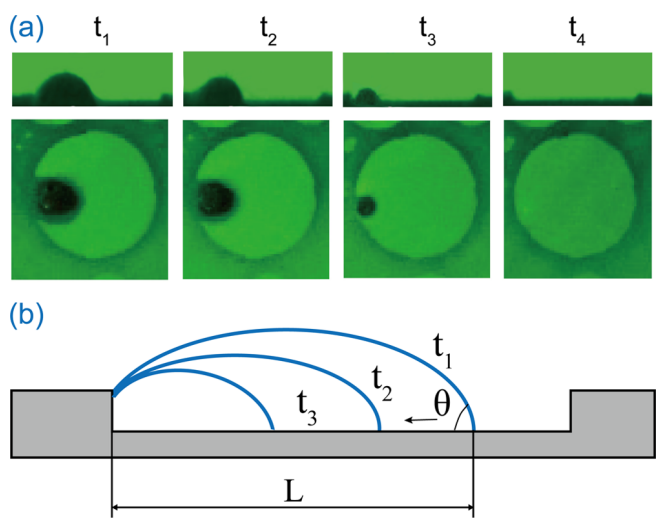

**Figure 4.** Shrinkage process of the entrapped microbubbles, in the CA-mode. (a) The side and bottom views of a shrinking microbubble. One side of the three-phase contact line of the microbubbles is pinned to the edge of the microwell. The other side recedes while the microbubble dissolves. (b) Cartoon of the microbubbles dissolving. During microbubble shrinking, it is pinned at one side, and both height and lateral diameter decrease with time.

To investigate how the contact angle changes during microbubble shrinkage, several individually entrapped microbubbles were tracked in time, using the three-dimensional confocal microscope. As an example, we consider here three entrapped microbubbles, as shown in Figure 5. Figure 5a–c are the side view images of the shrinking microbubble at different times. All the three microbubbles exhibit similar dissolving behavior, with one side being pinned at the edge of the wall and the other side retracting with time. During the process, one can clearly see that the contact angles of the individual microbubbles remain almost constant, as shown in Figure 5d, which implies that the microbubbles dissolve in the constant contact angle mode.

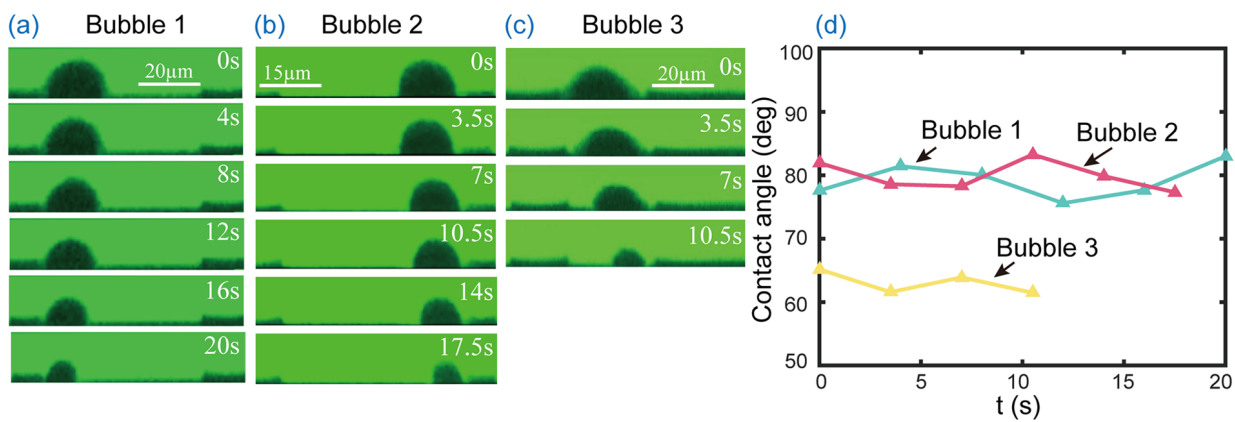

**Figure 5.** (a–c) Side view images of an individual dissolving microbubble at different times. The three microbubbles behave the same, with one side of the three-phase contact line pinned at the edge of the wall while the other side keeps receding with time. (d) Change of the contact angle of the three microbubbles with time. During shrinkage, the contact angles remain almost constant.

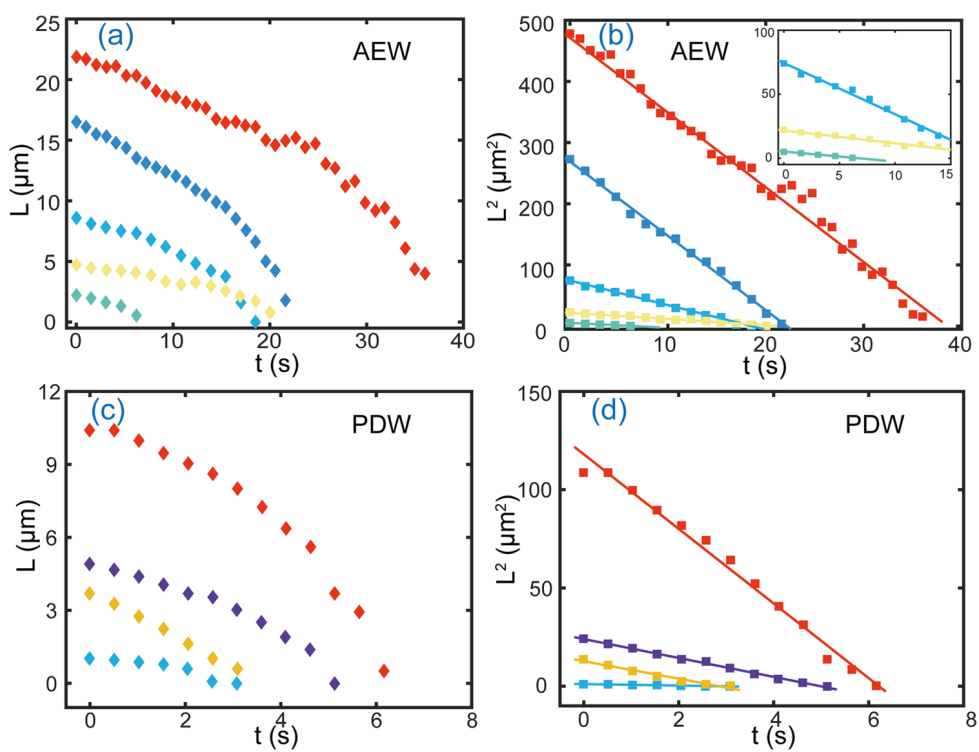

**Figure 6.** Change of the microbubble lateral diameter $L$ (a, c) and its square $L^2$ (b, d) as a function of time during microbubble shrinkage in air equilibrated water (a, b) and partially degassed water (c, d), respectively. In both cases, $L^2$ decreases linearly with time.

Under the conditions of constant contact angle shrinkage, on a plane surface spherical-cap-shaped microbubbles dissolve through air diffusion from microbubble to liquid, governed by[40]

$$L^2(t) = L_0^2 - \kappa(\zeta)t \qquad (1)$$

In the equation, $\kappa$ is the dissolution rate and is given by

$$\kappa = \frac{8Dc_s\zeta}{\rho} \frac{f(\theta)}{3g(\theta)} \qquad (2)$$

where $D$ is the diffusion coefficient, $c_s$ is the solubility of air in water, $\rho$ is the gas density in the microbubble, $\theta$ again is the contact angle of microbubble from the gas side, and $\zeta$ is the air undersaturation of water defined as $\zeta = 1 - c_\infty/c_s$, and

$$f(\theta) = \frac{\sin\theta}{1+\cos\theta} + 4\int_0^{+\infty} \frac{1 + \cosh 2\theta\zeta}{\sin 2\pi\zeta} \tanh((\pi - \theta)\zeta)\, d\zeta \qquad (3)$$

and

$$g(\theta) = \frac{\cos^3\theta - 3\cos\theta + 2}{3\sin^3\theta} \qquad (4)$$

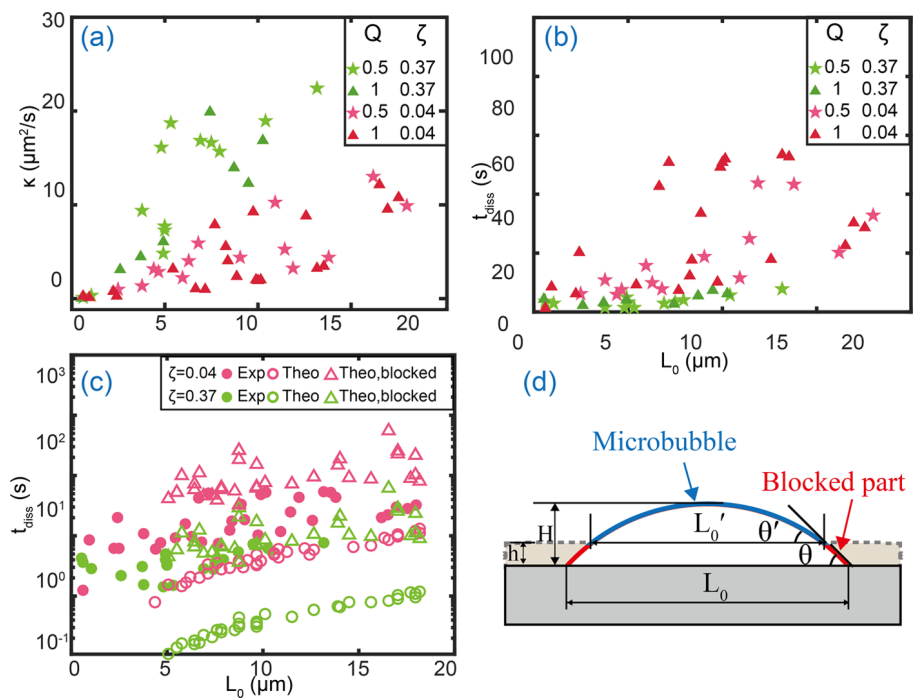

**Figure 7.** Comparison of the dissolution rate $\kappa$ and dissolution time $t_{diss}$ for microbubbles in AEW and PDW for two different flow rates $Q$ (value given in mL/min). (a) The value of $\kappa$ in PDW is higher than that in AEW. (b) The dissolution time $t_{diss}$ in AEW is larger than that in PDW. Clearly, the microbubbles dissolve faster in the water with a higher undersaturation value, independent of the flow rates. (c) The experimental dissolution time $t_{diss}$ is larger than the theoretical one for a spherical-cap-shaped bubble on a plane surface, both in AEW and PDW. The dissolution of microbubbles in the experiments is delayed, due to the wall blockage effect of the gas diffusion out of the entrapped bubble. This was supported by a quantitative estimation of the dissolution time (eqs 9 and 10) when the bottom part of the bubbles is blocked (data points correspond to the upper open triangles). (d) The geometry and notation for the microbubble with the bottom part blocked by the side wall of the microwell.

For an air–water solution with a certain undersaturation value, the dissolution rate $\kappa$ only depends on the contact angle $\theta$. Therefore, a constant contact angle leads to constant $\kappa$. As a result, $L^2$ is expected to linearly decrease with $t$, see eq 1. To verify the linear dependence, the lateral diameter $L(t)$ of the entrapped microbubbles was first tracked in air equilibrated water, as shown in Figure 6a. The value of $L^2(t)$ is plotted in Figure 6b. Indeed, $L^2(t)$ linearly decreases with time, which is consistent with the constant contact angle mode, and a spherical-cap bubble on a plane surface. Note that this latter assumption is not given, as the bubbles in our experiments are not sitting on a plane surface, but in a microwell, where the diffusive gas flux is partially blocked. Therefore, quantitatively the dissolution rate should be slightly slower than that for a bubble on a plane surface.

The constant contact angle dissolution mode, as well as the linear dependence of $L^2$ on $t$ holds for all gas concentrations. To verify this, we conducted further experiment in partially degassed water. The results are shown in Figure 6c,d, showing that $L^2$ still linearly decreases with time.

We now become more quantitative: from eq 2, which holds for the spherical-cap bubble on a plane surface, one can see how the value of the dissolution rate $\kappa$ depends on the undersaturation level $\zeta$ of water. This implies that the dissolution time $t_{diss}$ (defined by $L(t) = 0$) of the microbubbles also depends on $\zeta$. The value of $t_{diss}$ can be obtained by rewriting eq 1 as

$$\frac{L^2(t)}{L_0^2} = 1 - \frac{t}{t_{diss}} \qquad (5)$$

where

$$t_{diss} = \frac{L_0^2}{\kappa} = \frac{L_0^2 \rho}{8 D c_s \zeta} \frac{3g(\theta)}{f(\theta)} \qquad (6)$$

We can see that in this purely diffusive regime and for a spherical-cap bubble on a plane surface, the dissolution time $t_{diss}$ quadratically depends on the initial lateral microbubble diameter $L_0$, and is inversely proportional to the undersaturation. The relative gas concentration in AEW and PDW used in this study is 0.96 and 0.63, respectively. The corresponding undersaturation $\zeta$ is 0.04 and 0.37. Two different flow rates $Q$ were applied in the experiment for both PDW and AEW cases. In Figure 7, the dissolution rate $\kappa$ and the dissolution time $t_{diss}$ are compared for microbubbles with various initial lateral diameters for the two undersaturation levels. The results in Figure 7a show that $\kappa$ in PDW is higher than that in AEW, which is consistent with eq 1. It means that the microbubbles dissolve faster in water with a higher $\zeta$. In addition, we observed that $\kappa$ varies even for the same lateral diameter and slightly increases with increasing $L_0$. This is believed to be due to different contact angle values even for microbubbles with similar sizes, due to the difference in pinning. In Figure 7b, we can see that the measured $t_{diss}$ in AEW is larger than that in PDW, which agrees with eq 5. From the results shown in Figure 7, we conclude that neither the dissolution rate nor the lifetime is affected by the flow rates.

Moreover, according to eq 6, the theoretical dissolution time for a spherical-cap-shaped bubble on a plane surface $t_{diss}$ can be obtained with the measured contact angles by taking $D = 2 \times 10^{-9}$ m$^2$/s, $c_s = 0.023$ kg/m$^3$, and $\rho = 1.169$ kg/m$^3$.[42] A comparison of the experimental and theoretical results of $t_{diss}$ is shown in Figure 7c. One can see that the experimental value of $t_{\text{diss}}$ in both AEW and PDW is obviously larger than that calculated with eq 6. This means that the actual lifetime of microbubbles is longer than that expected from the diffusion theory for a bubble on a plane surface.

There are three possible factors which can be responsible for the discrepancy: the partial blockage of the gas outflux by the well walls, collective effects of neighboring microbubbles, and addition partial blockage by absorbing dye.[41−44] Since the bubble is not spherical-cap-shaped, sitting on a plane surface, but trapped in a microwell, the presence of the microwell side wall will lead to a partial blockage of the diffusive gas outflux. On the one hand, the edge of the microwell serves as anchors where the contact line is pinned. At this pinned portion of the microbubble, the side wall of the microwell blocks the diffusive outflux. Compared to micro-bubbles at flat sample surfaces, the gas cannot escape from the side wall of the microwell. This helps to slow down the bubble dissolution. On the other hand, even for the portion of the detached three-phase contact line from the opening of the microwell, the gas still cannot escape freely because the side wall of the microwell and the deepening partially block the gas diffusion. Both geometric effects of the microwell on the diffusive flux contribute to the increased $t_{\text{diss}}$.

We now roughly estimate how large this diffusive blockage effect can be. We assume that the bubble is partially blocked with the well wall of $h = 1 \ \mu m$ (the well depth on our sample is $0.8-1.2 \ \mu m$), as depicted in Figure 7d, where the lower part of the bubble is completely blocked by the side wall. The initial lateral diameter $L_0$, the height $H$ (with the bottom side of the well as a reference), and the contact angle $\theta$ of the partially blocked bubble are obtained from experiment. The corresponding parameters $L_0'$ and $\theta'$ for the unblocked part of the bubble are given as (see Figure 7d)

$$\theta' = \arccos \frac{L_0^2 - 4H^2 - 8Hh}{L_0^2 + 4H^2} \qquad (7)$$

$$L_0' = \frac{L_0^2 + 4H^2}{4H} \sin \theta' \qquad (8)$$

The mass loss $dM/dt$ of the partially blocked bubble then is[40,45]

$$dM(L', \theta')/dt = -\frac{\pi}{2} L' D c_s \zeta f(\theta') \qquad (9)$$

The mass $M$ of the bubble is

$$M(L, \theta) = \rho \frac{\pi}{8} L^3 g(\theta) \qquad (10)$$

According to eqs 9 and 10, and the geometric relationship in eqs 7 and 8, we can numerically determine the time when $M(L, \theta) = 0$ for the constant contact angle $\theta$. In this way, the theoretical dissolution time $t_{\text{diss}}$ in the partially blocked case is acquired, as shown in Figure 7c (open triangles). It is clear from the comparison that the effect of partial blocking is substantial. Note that theoretical dissolution time $t_{\text{diss}}$ in the blocked case is slightly larger than the experimentally measured ones. We expect that the real blockage effect should be smaller than what we have estimated here, because the bubbles are only partially pinned at the wall. In addition, the specific pinning portion along the three-phase contact line varies for different bubbles and changes during the dynamic process. As a result, it is difficult to get a completely quantitative evaluation of the partial blockage effects.

Regarding the effect of the neighboring microbubbles on the dissolution process, the distance between neighboring micro-bubbles is crucial. On our sample, the shortest distance between the neighboring bubbles is about 16 $\mu m$, which for most bubbles is larger than their diameter, see Figure 7. In one of our recent works,[41] we have calculated the effect of neighboring droplets (which have the same diffusive dynamics) on the dissolution time. For the closest packaging of droplets there is with a distance of 5 $\mu m$, on a footprint diameter of 10 $\mu m$. This is much closer than what we have here. Nonetheless, the dissolution time only increases by 60% when increasing the microwell distance from 5 $\mu m$ (where there are collective effects) to 20 $\mu m$ (hardly any collective effects). This is much less than the above blockage effect.

Finally, we discuss the effect of the addition of dye, which indeed will result in a slight increase of the dissolution time. The reasons are 2-fold: (i) the dye as a surfactant will slightly lower the surface tension and (ii) more importantly, the dye attachment to the interface will lead to a partial blockage of diffusion through the interface. These two effects in principle would contribute to the larger measured dissolution time, compared to the theoretical prediction. However, to minimize these dye effects, we used a relatively low dye concentration of 2 mg/mL. For such low concentrations, the dye effect on the surface tension is minimal: it decreases from 72.1 to 71.3 mJ/m² with the concentration of the dye from 0.1 to 20 mg/mL.[46] In summary, it is clear that the partial blockage of the gas outflux through the air−water interface is the main reason for delayed dissolution, due to geometric constrains by the walls of the well and to some degree also due to the partial coverage of the interface by the dye.

## ■ CONCLUSIONS

In this study, the temporal evolution of microbubbles at solid−liquid interfaces of immersed structured PS surfaces was systematically investigated. The results clearly show that the microwells on hydrophobic surfaces are able to trap gas and form microbubbles. The microwells were initially fully covered by the entrapped microbubbles that are pinned at the edges of microwells. Then part of three-phase contact line detaches from the microwell edges and the microbubbles rapidly shrink to a smaller size. Subsequently, the microbubbles undergo shrinkage in a constant contact angle mode due to gas diffusion into the liquid. Experimental results show that the square of footprint area of the microbubbles decreases linearly with time, which further confirms the constant contact angle dissolution mode. In addition, our results show that a higher under-saturation enhances microbubble dissolution, while the flow rates remarkably have no influence on the dynamics of microbubble entrapment and dissolution. We also see that geometric blockage effects due to the microwell lead to reduced dissolution rates compared to free bubbles on plane surfaces.

In general, we have shown that the position and size of interfacial microbubbles can be controlled. It is clear that the surface microstructures lead to gas entrapment. The amount of the entrapped gas directly depends on the size of the surface structures. In addition, the quantitative studies under various undersaturation and flow rate conditions give a rough estimation on the microbubbles lifetime and stability. We believe that this work will lead to a better understanding of the mechanism of interfacial nanobubble formation and provide an effective way to more stable and reproducible interfacial micro- and nanobubble formation.

## ■ AUTHOR INFORMATION

**Corresponding Authors**
*E-mail: wangyuliang@buaa.edu.cn (Y.W.).
*E-mail: xuehua.zhang@ualberta.ca (X.Z.).
*E-mail: d.lohse@utwente.nl (D.L.).

**ORCID**
Binglin Zeng: 0000-0001-8729-1944
Xuehua Zhang: 0000-0001-6093-5324
Detlef Lohse: 0000-0003-4138-2255

**Notes**
The authors declare no competing financial interest.

## ■ ACKNOWLEDGMENTS

This work is supported by the National Natural Science Foundation of China (Grant No. 51775028) and Beijing Natural Science Foundation (Grant No. 3182022). The authors thank the Dutch Organization for Research (NWO) and the Netherlands Center for Multiscale Catalytic Energy Conversion (MCEC) for financial support. D.L. also acknowledges financial support by an ERC—Advanced Grant and by NWO—CW.